\documentclass[runningheads]{llncs}
\usepackage{graphicx}
\usepackage{multirow}
\usepackage{adjustbox}
\usepackage{lipsum}
\usepackage{algorithmic}
\usepackage{graphicx}
\usepackage{float}
\usepackage{url}
\usepackage{colortbl}
\usepackage{tcolorbox}

\begin{document}

\title{Kuksa: A Cloud-Native Architecture for Enabling Continuous Delivery in the Automotive Domain}

\titlerunning{Kuksa: a cloud-native architecture}

\author{Ahmad Banijamali\inst{1}\orcidID{0000-0002-6283-142X} \and
Pooyan Jamshidi\inst{2}\orcidID{0000-0002-9342-0703} \and
Pasi Kuvaja\inst{1}\orcidID{[0000-0002-1488-6928} \and
Markku Oivo\inst{1}\orcidID{0000-0002-1698-2323}}
\authorrunning{Banijamali et al.}
\institute{M3S Research Unit, ITEE Faculty, University of Oulu, Finland \email{\{firstname.lastname\}@oulu.fi}\\ \and
Computer Science and Engineering Department, University of South Carolina, USA\\
\email{pjamshid@cse.sc.edu}}
\maketitle
\vspace{-.5cm}
\begin{abstract}
Connecting vehicles to cloud platforms has enabled innovative business scenarios while raising new quality concerns, such as reliability and scalability, which must be addressed by research.\ Cloud-native architectures based on microservices are a recent approach to enable continuous delivery and to improve service reliability and scalability.\ We propose an approach for restructuring cloud platform architectures in the automotive domain into a microservices architecture.\ To this end, we adopted and implemented microservices patterns from literature to design the cloud-native automotive architecture and conducted a laboratory experiment to evaluate the reliability and scalability of microservices in the context of a real-world project in the automotive domain called Eclipse Kuksa.\ Findings indicated that the proposed architecture could handle the continuous software delivery over-the-air by sending automatic control messages to a vehicular setting.\ Different patterns enabled us to make changes or interrupt services without extending the impact to others.\ The results of this study provide evidences that microservices are a potential design solution when dealing with service failures and high payload on cloud-based services in the automotive domain.

\end{abstract}
\vspace{-1cm}
\section{Introduction}

\begin{figure}[t]
\centering
\scalebox{0.45}{\includegraphics{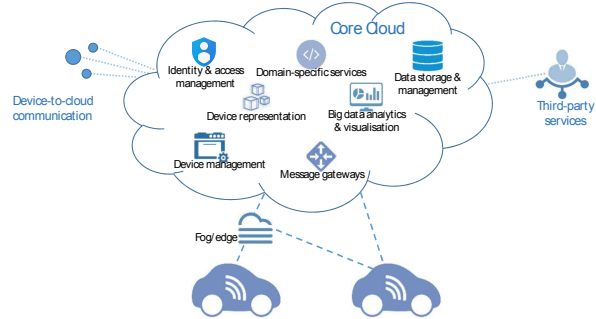}}
\vspace{-3cm}
\caption{Cloud platforms in the automotive domain}
\label{fig1}
\vspace{-.7cm}
\end{figure}

In recent years, there has been an increased focus from industry and academia to investigate cloud platform architectures that enable continuous software delivery (CD) in vehicles \cite{ebert2017automoitve}.\ Many industries have started to look for CD solutions as they need to release quality software more frequently, better respond to automotive market changes, avoid vehicle recalls, improve productivity, and increase customer satisfaction \cite{shavit2007firmware}.\ For this purpose, vehicular software and information resources are being virtualised and designed as services in the cloud \cite{he2014developing}.\ Cloud platforms in the automotive domain (ACPs) provide the possibilities to exchange data beyond vehicles \cite{armbrust2010view}, connect vehicles to other objects in the environment, update automotive software using wireless communications systems (over-the-air) \cite{zeller2013towards}, and enable many more business services in the cloud (Figure \ref{fig1}).

Nevertheless, the migration of software delivery to ACPs has raised new research challenges.\ For example, vehicle-to-cloud (V2C) data transmission requires low latency and high reliability to satisfy the requirements of real-time systems \cite{lu2014connected}.\ Scalability is another challenge that demands the decomposition of functionalities and efficient data management \cite{haberle2015connected}.\ Furthermore, the resiliency configuration explains runtime behaviour and faulty components \cite{haberle2015connected}, and security is a major requirement for protecting vehicles from malicious attacks \cite{zhang2014defending}.

In addition, as for the migration process towards distributed systems, such as cloud-native architectures, many architecture designs fail as long as their goal is to only replace the existing legacy architecture with a virtualised environment in the cloud \cite{balalaie2015migrating}.\ The reasons may include but are not limited to a lack of solid business cases for cloud migration, neglecting adequate support teams, migrating at once to the cloud, and not considering applications' architecture refactoring \cite{balalaie2018microservices,bass2015devops}.\ Consequently, the benefits from migration to the cloud platforms could be trivial, as the failure can happen anytime \cite{balalaie2015migrating}.

Despite the importance of CD and the mentioned quality challenges in ACPs, there has been insufficient focus from research that provides practical insights into designing software architectures that address those quality concerns \cite{Chen-microservice}.\ Due to the impact of microservices on cloud-native architectures with respect to quality requirements, such as reliability, scalability, availability, and fault-tolerance \cite{balalaie2018microservices,bass2015devops}, microservices can be a potential solution for the existing challenges in ACPs.\ In relation to this, the ultimate objective of our paper is to investigate whether microservices \textit{can enable over-the-air (OTA) continuous delivery in ACPs while improving reliability and scalability in this domain}.\ We have proposed a microservices architecture based on a real-world project in the automotive domain called Eclipse Kuksa and conducted a laboratory experiment to evaluate the architecture with respect to the mentioned quality attributes.

The results of this study can benefit industrial practitioners and academic researchers in the domains of automotive software engineering and cloud platform design.\ The study is aimed at researchers who would like to gain insight into the application of microservices in the domain of ACPs.\ From the practitioners' perspective, the findings provide experimental results for the reliability and scalability of microservices in a real-world industrial case in the automotive domain.\ The key contributions of the study are: (1) assessing the relative extent to which cloud-native architecture can enable continuous delivery in the automotive domain and (2) evaluating the role of microservices patterns in improving the reliability and scalability of services in this context.
\vspace{-.3cm}

\section{Background}
\label{II}
\vspace{-.2cm}

\subsection{Microservices}
\vspace{-.1cm}

Monolithic architectures are usually successful when the whole system is small and the number of functions is low \cite{dragoni2017microservices}.\ Increasingly, the number of end users requires more deployment in the cloud \cite{dragoni2017microservices}, as every time that we apply a change to a small part of an application, we need to build and deploy the whole monolithic system again \cite{balalaie2015migrating}.\ Furthermore, scalability means scaling the whole application rather than a part of the components that requires more resources \cite{MartinFowler}.\ As a consequence, many companies, such as Netflix, Amazon, and Atlassian, have migrated to more scalable and reliable architectures like microservices.

As for distributed systems, microservices are used to design fine-grained, modular services that have different life cycles but work together \cite{newman2015building}.\ Each service deploys independently \cite{balalaie2016microservices} using a potentially different deployment framework typically in the cloud \cite{pahl2016microservices}, scales independently \cite{thones2015microservices}, is tested individually, and accomplishes responsibilities independently \cite{thones2015microservices} while communicating through lightweight mechanisms, such as RESTful APIs \cite{MartinFowler}.\ The relevant architecture breaks down a system into services, each as a business capability \cite{MartinFowler}.

Microservices promote a DevOps philosophy about separated small teams working together to meet the objectives of a large mission-critical system \cite{bass2015devops}.\ On the other hand, DevOps provides the framework for developing, deploying, and managing the microservices container ecosystem \cite{ebert2016devops}.\ In this architecture, a microservice is developed and maintained by one small team while coordination among the teams is minimised \cite{zhu2016devops}.\ It is noted that the largest size of the teams usually follows Amazon's notion of the ``Two Pizza Team'', meaning not a large group of people \cite{MartinFowler}.

Despite all the advantages that microservices bring to the architecture designs, they have several challenges that should carefully be addressed.\ For example, replacing a monolithic architecture with a large number of inter-connected microservices can increase latency and other performance issues \cite{bass2015devops}.\ Having a system that is currently being used in production, it is necessary to make the migration incrementally \cite{zhu2016devops} without data loss and interruption \cite{bass2015devops}, during which we need adequate frameworks and experience in how to proceed \cite{zhu2016devops}.\ Eventually, inconsistencies among microservices is another relevant challenge \cite{MartinFowler}.
\vspace{-.5cm}
\subsection{Software Architectures of Automoitve Cloud Platforms}
\vspace{-.1cm}

Convergence of the internet of things and cloud computing has enabled innovative business use cases, ecosystems, and players in the automotive domain \cite{he2014developing}.\ ACPs' application includes but is not limited to advanced vehicle connectivity, infotainment applications, voice and video data streaming, fleet management services, remote diagnostics and maintenance, and telematics services \cite{googlecloud,shiftmobility}.

Due to the increasing number of connected vehicles, the security, reliability, availability, robustness, and scalability of services are becoming new quality requirements in ACPs \cite{Haghighatkhah}.\ The extent of architectures in ACPs ranges from multi-layered architectures \cite{contreras2017internet} to service-oriented architectures (SOA) \cite{mietzner2011horizontal,yang-intelligent}.\ Datta et al.\ \cite{datta2015onem2m} designed a framework for connected vehicles to offer consumer-centric services and a uniform mechanism for describing and collecting vehicular sensors' data.\ The designed architecture applies technologies such as road side units (RSUs) and machine-to-machine (M2M) gateways, including the fog computing platform \cite{datta2015onem2m}.\ The authors argued that using fog computing technologies can improve the fault tolerance, reliability, and scalability of the system \cite{datta2015onem2m}.\ Scalability and interoperability have been addressed in another study \cite{rufino2017monitoring} in a modular architecture built upon DevOps practices to enable vehicle-to-everything (V2X) applications.\ The authors divided real-time applications for managing traffic into small modules to validate the functionality of the architecture \cite{rufino2017monitoring}.

A scalable and fault-tolerant data-processing design for real-time traffic-based routing was proposed by another study \cite{Serrano-realtime}.\ It argued that the designed architecture can serve a wide range of workloads and use cases with low-latency requirements \cite{Serrano-realtime}.\ Real-world scenarios of intelligent traffic system applications demonstrated the need for scalable big data analysis, service encapsulation, dynamic configuration, and optimisation strategies in this context \cite{Fiosina-bigdata}.\ Due to the technological variety in ACPs, architecture designs must assure stakeholders \cite{bass2015devops} that provisional services will meet the quality requirements at a specific level of cost and risk that is enforced by service level agreements (SLAs) \cite{o2007quality}.

\vspace{-.3cm}
\section{Research Questions and Method}
\label{III}
\vspace{-.2cm}
This section describes the study's objective, research questions, and research method.

\vspace{-.2cm}
\subsection{Objective and Research Questions}
\vspace{-.1cm}

The main objective of our study was to evaluate whether microservices can address CD in the context of ACPs and whether they can improve the reliability and scalability of services in this context.\ The research questions (RQs) for this study were as follows:

\begin{tcolorbox}
\begin{itemize}
\item RQ1: Can the microservices architecture design enable over-the-air continuous delivery from cloud platforms in the automotive domain?
\item RQ2: How can the microservices architecture design improve the reliability and scalability of services in cloud platforms in the automotive domain?
\end{itemize}
\end{tcolorbox}
\vspace{-.5cm}

\subsection{Research Method}
\label{research-method}
\vspace{-.1cm}

To design the target microservices architecture, we adopted a software architecture from a real-world project in the context of ACPs called Eclipse Kuksa (see Section \ref{IV}).\ It was important to initiate the migration process based on an existing project to review how the new architecture design could improve reliability and scalability in this domain.\ For the migration and refactoring process of the current architecture of Eclipse Kuksa, we applied microservices patterns from literature (e.g., \cite{balalaie2018microservices}).\ Each refactoring represented a small and controlled change, so it was possible to identify how the quality attributes changed.\ The codes are available on GitHub\footnote{https://github.com/ahmadbanijamali/Rover-Control-Experiment.git}.

Recent research \cite{stol2018abc} has explained that although it is critical to evaluate the requirements of a new software system to ensure system acceptance by users, real context evaluations are often complex.\ Before operating newly designed systems in real dynamic and complex environments, it is reasonable to assess them in laboratory setting experiments \cite{stol2018abc}.\ Thus, to evaluate the designed microservices architecture, we used laboratory experiments as the research method to answer the RQs of this study.
 
To date, there are several domain-specific services designed in Eclipse Kuksa.\ Among them, this study selected a service that is used for the purpose of motion control.\ Previous studies \cite{balalaie2018microservices,bass2015devops,levcovitz2016towards,newman2015building,taibi2018architectural} have proposed frameworks and parameters in which architecture designers select microservices for migration, for example, according to their value to end users (e.g., improved user experience regarding the availability of services) or the project organisation (i.e., information exchange scalability and resiliency support) \cite{bass2015devops}.\ We selected the \textit{motion control service} because of its value to end users and applicability in different scenarios.\ Furthermore, it demonstrates how end users can send control commands to vehicles from the cloud platform in Eclipse Kuksa using different user interfaces.\ It is a general service that can be part of many scenarios in this domain.\ The primary business driver for this service is to demonstrate OTA updates and messaging from the cloud to vehicles.\ This creates suitable grounds for future studies, e.g., on driver behaviour optimisation, natural language processing in vehicles, or OTA driver authentication.

Section \ref{V} provides more details of our evaluation setting and the technology stacks used in our experiment.

\vspace{-.5cm}
\section{Eclipse Kuksa}
\label{IV}
\vspace{-.3cm}

The Eclipse Kuksa\footnote{https://projects.eclipse.org/projects/iot.kuksa} utilises open, vehicle-independent protocols, ensuring lifetime value for vehicles through upgradable applications.\ It addresses application systems, software solutions, and services for the mass differentiation of vehicles.\ The ecosystem of Eclipse Kuksa is comprised of three main platforms, including the (1) in-vehicle platform, (2) cloud platform, and (3) an app IDE.\ The Eclipse Kuksa is supported by a wide range of integrated open source software technologies and development environments, such as automotive grade Linux (AGL) and Eclipse Paho for the in-vehicle platform and Eclipse-Hono, Eclipse Hawkbit, Eclipse MosQuitto, Keycloak, and InfluxDB in the cloud back-end.

\subsection{The Existing Architecture of Eclipse Kuksa}
\label{exisitngArch}

\begin{figure}
\centering
\scalebox{0.56}{\includegraphics{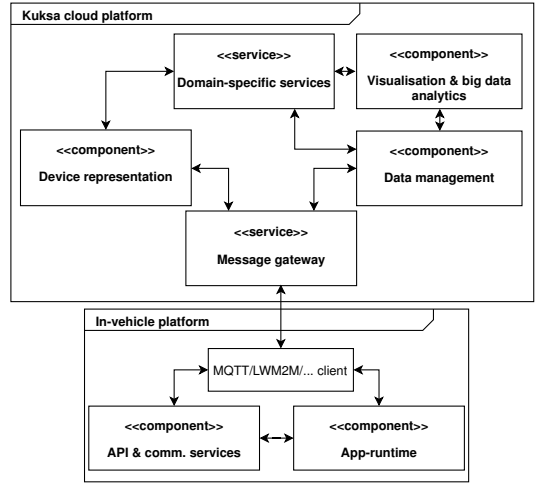}}
\vspace{-.2cm}
\caption{Software architecture of Eclipse Kuksa}
\label{fig2}
\vspace{-.6cm}
\end{figure}

Figure \ref{fig2} shows the components and services in the Eclipse Kuksa architecture.\ The architecture only provides information about the necessary components and services that we needed in our experiment in the scope of this paper.\ It neglects other parts of Eclipse Kuksa ecosystem, such as device management and representation, authentication and authorisation, and the app store.

\vspace{-.5cm}

\textit{Message Gateway.} The Eclipse Kuksa cloud platform (EKCP) sends and receives different types of messages from and to various sources, such as vehicles, devices, and third-party services.\ In general, messages include ``telemetry messages'' that depict data stemming from vehicles, devices, and sensors and ``commands and controls messages'' that are dedicated to the vehicles and device management components.\ The message gateway provides remote service interfaces for connecting vehicles and devices to the cloud back-end.

\textit{Data Storage and Management.} An important part of the realisation of the EKCP is the storage and management of vehicles' and IoT devices' data in the appropriate database management system (DBMS).\ Although data management is a central aspect of every cloud platform architecture, due to the wide range of vehicles and devices connected to ACPs, it is necessary to establish a well-defined data management system that can handle complexities related to big data, consistency, performance, scalability, and security.

\textit{Visualisation and Big Data Analytics.} The advances in the digitisation of the automotive domain have created a large amount of heterogeneous data coming from various sources.\ This has also yielded new requirements in terms of volume, variety, and velocity that are commonly called big data.\ The EKCP includes components and services to visualise and manage the big data in this domain.

\textit{Device Representation.} To realise the distinct functionality of domain-specific services, a digital representation is important.\ Digital twin offers the possibility to access and alter the state of a vehicle's functionality in a controlled manner.

\textit{Domain-specific Services.} The domain-specific services are developed according to different use cases and business scenarios on top of the in-vehicle platform.\ They can handle different functions and tasks in vehicles and beyond them.

\textit{In-vehicle Platform.} The communication protocols such as MQTT and LWM2M have enabled sending different messages from vehicles to the cloud and vice versa.\ The in-vehicle platform in Eclipse Kuksa includes an app runtime environment that is connected to an in-vehicle gateway, enabling software delivery and deployment in vehicles.

\vspace{-.3cm}
\subsection{The Proposed Microservices Architecture for the Eclipse Kuksa Cloud Platform}
\vspace{-.2cm}

\begin{figure}[H]
\centering
\scalebox{0.52}{\includegraphics{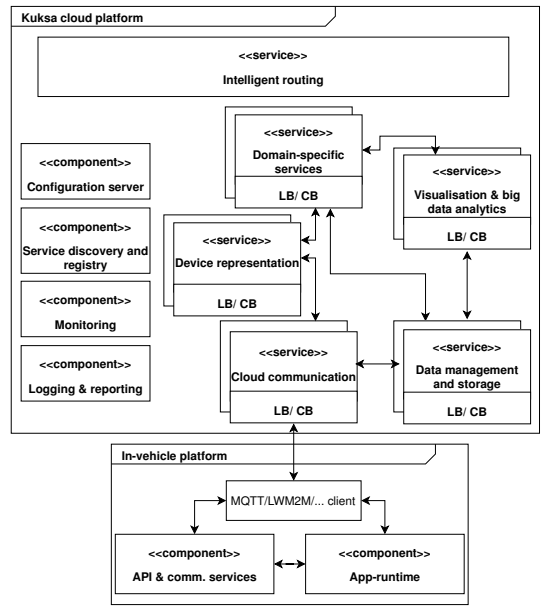}}
\vspace{-.4cm}
\caption{The microservices architecture in Eclipse Kuksa}
\label{fig3}
\vspace{-.7cm}
\end{figure}

Connected vehicles have high demands on the exchange of data between vehicles and a variety of services in the cloud.\ Due to the importance of the domain-specific services in ACPs, we selected a sample telemetry service that communicates with vehicles through sending command and control messages to vehicles (see Section \ref{research-method}).\ Figure \ref{fig3} shows our proposal for the refactored architecture of EKCP that is described in greater detail in this section.

\vspace{-.7cm}

\begin{figure}
\centering
\scalebox{0.4}{\includegraphics{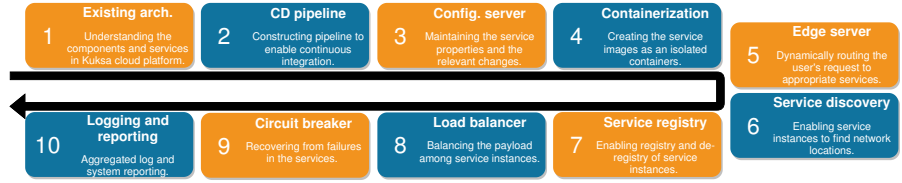}}
\vspace{-.3cm}
\caption{The migration process to a microservices architecture}
\label{fig4}
\vspace{-.7cm}
\end{figure}

The migration to a microservices architecture in EKCP is a step-by-step process including new components and modules and modifying the existing components (Figure \ref{fig4}).\ We started the process by creating a better understanding of the existing architecture (Section \ref{exisitngArch}) and introducing the CD pipeline.

\textit{Configuration Server.} According to previous research \cite{balalaie2018microservices}, we required two individual and separate repositories as source code storage and software configurations storage.\ The configuration server is a central place to support the externalised configuration and changes without rebuilding or restarting the services.\ The Spring cloud configuration server is a potential technology that stores each microservice property based on the service-ID.\ The properties can be stored in the cloud or in other repositories, such as in GitHub.

\textit{Containerisation.} The next step before establishing an intelligent routing (edge server) component was the containerisation of each service.\ This step is a part of the CD pipeline for building the container image for each service.\ The Docker and Docker Hub are the technology stacks used for this purpose.

\textit{Intelligent Routing (Edge Server).} This is the layer right after the user interface (UI).\ Edge server dynamically routes requests to the appropriate microservices.\ Thus, it is possible here to monitor the service usage, as all requests pass this layer.\ As an instance of the technology stack, Netflix provides Zuul as the front door for all requests from devices and web sites to the back-end.

\textit{Service Discovery.} Service instances dynamically find network locations of a service provider, which is critical for the service's auto-scaling and failures.

\textit{Service Registry.} In addition to service discovery, service registry registers and de-registers service instances.\ It stores addresses of each service as the service initiates and removes the addresses once it does not receive the heartbeat or the service is terminated.\ Spring Eureka provides the technology stack for service discovery and registry.

\textit{Load Balancer.} A purpose for migrating to a microservices architecture is to improve the scalability of each service based on the payload \cite{bass2015devops}.\ We used load balancers to distribute the payload among multiple instances of our services.\ Netflix Ribbon and Apache Zookeeper are examples of relevant technology stack.

\textit{Circuit Breaker.} Once the number of consecutive failures in services crosses a specific threshold (open state), we call the circuit breaker to either invoke a response code or return the latest cached data from the service provider.\ Once the timeout expires, the circuit breaker allows a limited number of test requests to service providers, and, if they pass, it changes to a closed state.\ Hystrix and NGINX are relevant technology stacks here.

\textit{Logging and Reporting.} To control what is happening in microservices, accessing the consolidated logs \cite{bass2015devops}, implementing infrastructure-level metrics, and creating a holistic view of the system, we need to establish an efficient logging and reporting functionality.\ The system is used for a variety of purposes, such as monitoring the traffic and service usages, identifying the cause of errors, and finding performance bottlenecks.\ Due to the wide scope, different technologies (i.e., Hystrix, Grafana, Kibana, and fluentd) are used for specific purposes.

\begin{figure}[H]
\centering
\scalebox{0.74}{\includegraphics{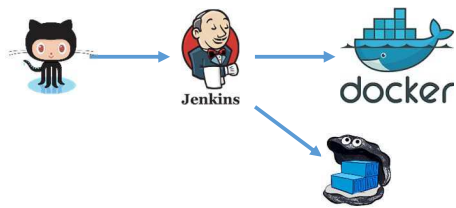}}
\vspace{-.2cm}
\caption{The continuous delivery pipeline}
\label{fig5}
\vspace{-.4cm}
\end{figure}

\textbf{Continuous Delivery Pipeline.}
To establish a CD pipeline, we required continuous integration using following components.\ Jenkins was the solution used as the continuous integration server to build and deploy the applications.\ Docker was the tool that we used for the containerisation of applications and to isolate them from each other.\ The Docker Hub, as the repository of Docker container images, pulls images from Docker's public registry instance.\ Figure \ref{fig5} shows the CD pipeline in EKCP.

\vspace{-.6cm}

\section{Evaluation}
\label{V}
\vspace{-.15cm}

\subsection{Experimental Setting}
\vspace{-.1cm}

To evaluate CD in the proposed architecture, we considered that our service sent automatically-generated updates as specific calls to forty vehicles in a specific region of the city.\ The calls were similar as they were demonstrating one released update.\ The software delivery cycle that the calls sent to the vehicles was one minute.\ In each call, we changed ``next move direction'' in the rover and the designed architecture should continue the message delivery without interruption.\ We ran the experiment for a duration of one hour to record how different microservices patterns behave in a CD environment in ACPs.\ We reviewed what percentages of calls is sent successfully to the rover and provide a statistics of successful and failed calls to show the CD performance in our design.

To review the scalability and reliability of the services in our designed architecture, we deployed three different scenarios.\ We aimed to measure metrics such as service downtime, recovery time, and load sharing behaviours.\ We registered four instances of Backserver service and one Client service (see Section \ref{V}) on a Spring Eureka server.\ The experimental scenarios were as follows.

\begin{enumerate}
\item During the first 10 minutes, all services were up and running.\ Half of the Backserver instances (two instances) shutdown automatically at 00:10 and re-started simultaneously at 00:15.
\item All service instances from the Backserver shut down automatically at 00:20 and started gradually (one by one) every five minutes until they all came up at 00:40.
\item All service instances from the Backserver went off at 00:45 and re-started simultaneously at 00:50.
\end{enumerate}

\begin{figure}[t]
\centering
\scalebox{0.43}{\includegraphics{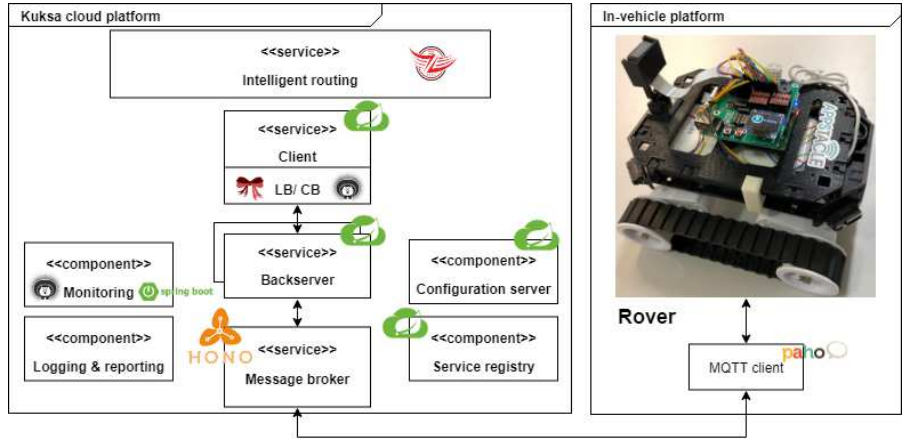}}
\vspace{-.3cm}
\caption{The experimental setting}
\label{fig6}
\vspace{-.5cm}
\end{figure}

Figure \ref{fig6} presents the experimental setting in this study, including the different services, components, and technology stacks.
\vspace{-.4cm}

\subsubsection{The cloud Back-end.}
We developed the Backserver and Client services using Spring Boot.\ All microservices were running on a computer with an Intel Core i7-6600U CPU @2.6 GHz and 20 GB installed RAM.\ Eclipse Hono version 7.0 was used as the message broker to connect the Backserver to the in-vehicle platform via MQTT using a 4G connection.\ The Hono instance was placed on an Azure Kubernetes service (AKS) cluster.

The Client service automatically triggered the delivery to the Backserver instances.\ Each message delivered to the rover contained the ``rover id'', ``speed control'', and ``next move direction''.\ The Backserver was responsible for sending the messages to the Hono instance and from there to the rover.\ The microservice patterns and technologies used are shown in Table \ref{microservices_pattern}.

\vspace{-.5cm}
\begin{table}[ht]
\centering
\caption{Microservice patterns and technologies used in this experiment}
\vspace{-.2cm}
\begin{tabular}{|l|l|p{6cm}|}
\hline
\rowcolor{lightgray}Pattern & Technology & Customised configuration\\
\hline
Intelligent routing & Netflix Zuul& serviceId: backserver, serviceId: Client\\
\hline
Load Balancing& Netflix Ribbon&Server list refresh interval: 2s\\
\hline
Circuit breaker& Hystrix&Sleep window: 5s, Request volume threshold: 20, Error threshold: 50\% \\ 
\hline
Configuration server & Eureka&--\\
\hline
Service registry& Eureka& eureka.client.register-with-eureka=false eureka.client.fetch-registry=false\\
\hline
Monitoring& Hystrix dashboard&--\\
\hline
\end{tabular}
\label{microservices_pattern}
\vspace{-1.2cm}
\end{table}

\subsubsection{The in-vehicle Platform.}
To demonstrate the outcomes of the experiment, we used a rover, which is an open source mobile robot.\ The rover includes a Raspberry Pi 3 Model B (RPi3), a motor driver layer (Arduino), and a RoverSense layer designed for in-vehicle communication demonstrations.\ A customised software (called roverapp\footnote{https://app4mc-rover.github.io/rover-app/}) was designed that runs on a Linux-based embedded single board computer (i.e., RPi3).\ The roverapp includes an API to handle various functions in the rover, such as motion control.

In addition to the commands sent to the rover, the RoverSense layer sends telemetry data from different sensors, such as infrared proximity sensors, ultrasonic sensors, temperature and humidity sensors, and an accelerometer to the cloud.\ The roverapp creates the possibility of real-time video streaming to the cloud platforms, such as Azure or AWS.\ It also allows the marker detection used in platooning or autonomous driving scenarios.

The rover's features' applications and tooling use AGL as the operating system, which runs on RPi3.\ The in-vehicle Kuksa layers, including a middleware layer (containing Kuksa APIs and Eclipse Paho) and an application layer (containing a runtime and sandbox environment), run on top of AGL.\ These two layers enable functions such as communication to the cloud via MQTT and third party applications' implementation.

\vspace{-.5cm}
\subsection{Results}
\label{VI}
\vspace{-.2cm}

This section is structured to address the research questions and includes the aggregated results of our experiment.
\vspace{-.5cm}
\subsubsection{RQ1.\ Can the microservices architecture design enable over-the-air continuous delivery from cloud platforms in the automotive domain?}

\begin{figure}[H]
\centering
\scalebox{0.57}{\includegraphics{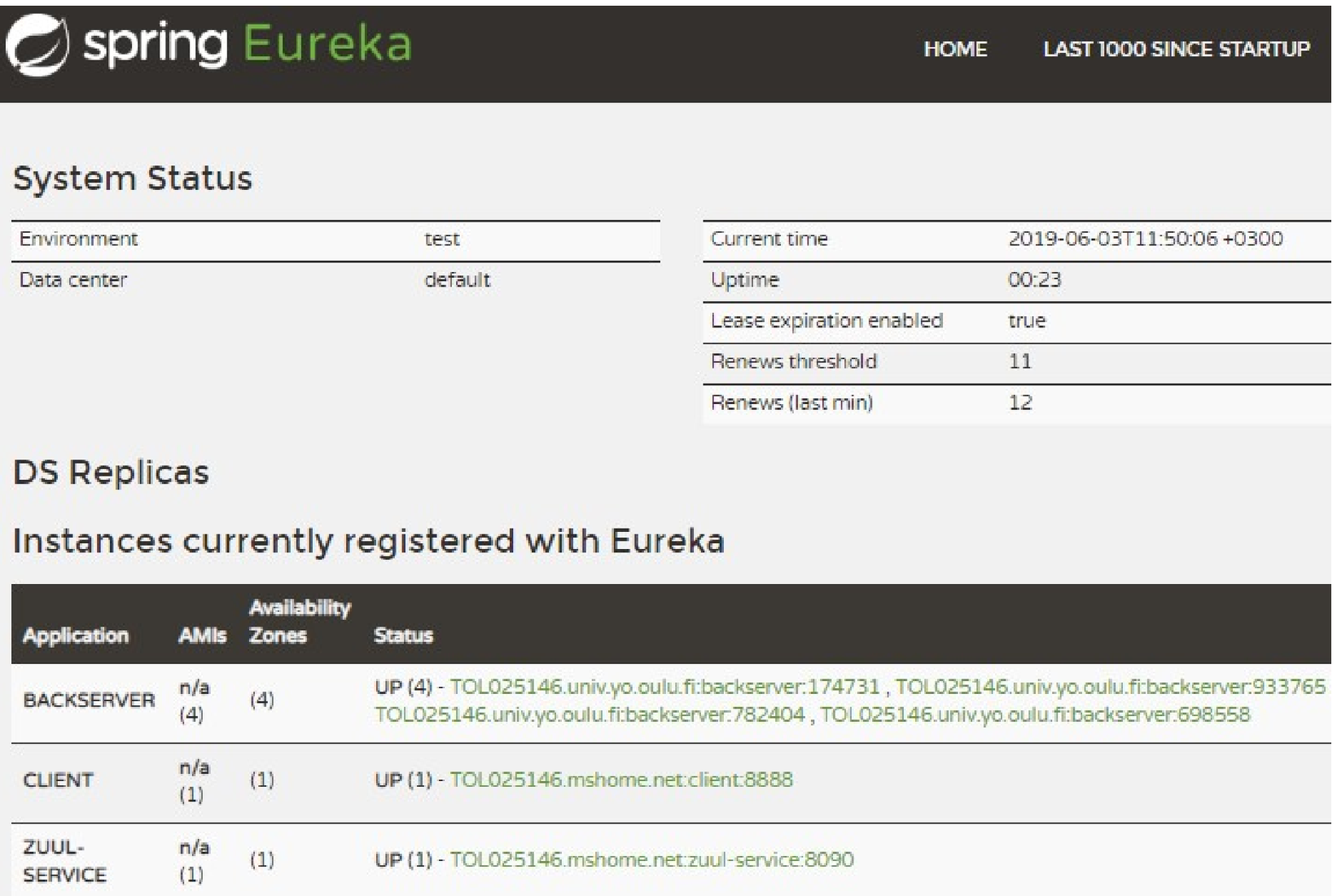}}
\vspace{-.2cm}
\caption{Registered services for the designed architecture}
\label{fig7}
\vspace{-.7cm}
\end{figure}

CD helps teams to produce applications in short cycles and ensures that the software can be reliably released at any time.\ Figure \ref{fig7} shows the service registry dashboard in a Spring Eureka server.\ It shows that four instances of the Backserver and one Client service were up and running at the time of the experiment.

\vspace{-.5cm}

Table \ref{CD-measure} shows the aggregated results of the duration that each service instance of the Backserver was up during our designed scenarios.\ In addition, it shows statistical information on the service resiliency in our setting.

\begin{table}
\centering
\caption{Experiment results using the designed microservices architecture}
\vspace{-.25cm}
\begin{tabular}{p{5cm}cp{4.5cm}c}
\hline
\rowcolor{lightgray}&& \textbf{No.\ of running instances in three scenarios} & \textbf{Time}\\
\hline
\multirow{3}{*}{Total duration of experiment}& \multirow{3}{*}{60 min.} & zero (shutdown all instances) & 10 min.\\\cline{3-4}
&&one& 5 min.\\\cline{3-4}
\multirow{3}{*}{Cycle time} & \multirow{3}{*}{1 min.} &two & 10 min.\\\cline{3-4}
&&three & 5 min.\\\cline{3-4}
&&four & 30 min.\\\hline
\hline
\rowcolor{lightgray}\textbf{Circuit breaker status} & \textbf{No.} & \textbf{Circuit breaker status} & \textbf{No.} \\
\hline
Success (execution completed with no errors) & 22 & Failure (execution threw an Exception) & 24\\
\hline
Timeout (execution started, but did not complete in the allowed time) & 4 & Short-Circuited & 0\\
\hline \hline
Quickest time to call Backserver when came up&&1 min.&\\
\hline
\end{tabular}
\label{CD-measure}
\vspace{-.65cm}
\end{table}

During our experiment, we could make changes (shutdown, re-start, and update the code) in a service without affecting other services.\ According to our experimental scenarios, Backserver instances set up and down multiple times, even though it did not impact other available services.\ It was easy to make changes on a service, e.g., updating the listening port or rover direction, without interruption to other services.

Our findings indicated that although we had a number of failed calls and timeout errors due to the following reason, the circuit breaker could prevent cascading failures to other services.\ The Client service talked with the service registry to receive the IP addresses of available Backserver instances and used its load balancer to choose one of them.\ The Client service could not know directly that a Backserver instance was no longer available.\ This is the job of the service registry to continuously discover which Backserver instances are dead or alive via heartbeat mechanisms.\ During our experiment, the Backserver instances shutdown several times while the Client service could not get the list of the remaining instances from the service registry in real-time.\ In this approach, the service discovery logic tightly coupled with clients, in which it could improve through other approaches, such as server-side service discovery.

\begin{tcolorbox}
\textbf{Summary.} The designed architecture preserved continuous software delivery by automatic registering and de-registering service instances and continuing OTA software delivery after each change.
\end{tcolorbox}

\vspace{-.65cm}
\subsubsection{RQ2.\ How can the microservices architecture design improve the reliability and scalability of services in cloud platforms in the automotive domain?}
Table \ref{CD-scalability} shows a summary of the results of the total calls on each Backserver instance.\ The Client service sent more calls on the Backserver instances that were up for a longer time in our scenarios.\ In total, we had 990 successful calls distributed among four Backserver instances to control the rover speed and movement direction.

\vspace{-.55cm}
\begin{table}[H]
\centering
\caption{The number of calls on the Backserver instances}
\vspace{-.2cm}
\begin{tabular}{p{4.5cm}p{1.5cm}p{2cm}p{1.5cm}}
\hline
\multirow{4}{*}{Total calls sent to rover}& \multirow{4}{*}{990}&Service \#1 &455\\\cline{3-4}
&&Service \#2 &276\\\cline{3-4}
&&Service \#3 &153\\\cline{3-4}
&&Service \#4 &106\\\hline
\end{tabular}
\label{CD-scalability}
\end{table}

\begin{figure}[ht]
\centering
\scalebox{0.43}{\includegraphics{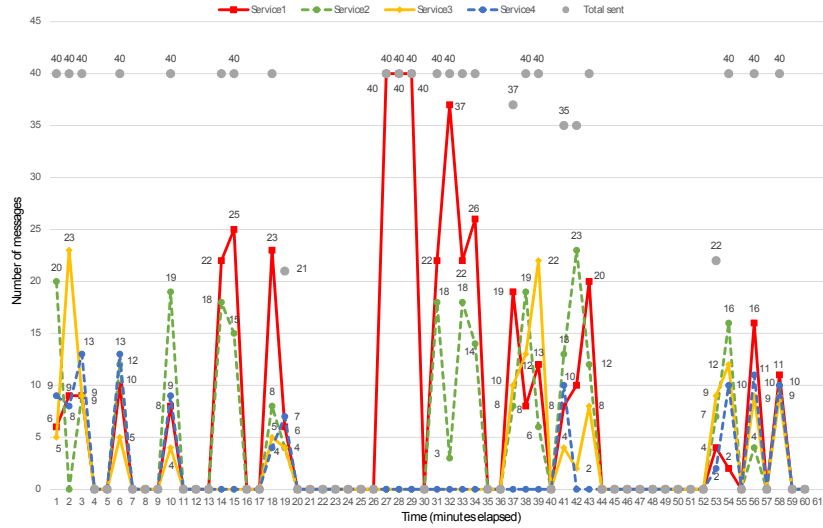}}
\vspace{-.4cm}
\caption{The number of calls on each service in three scenarios}
\label{fig8}
\vspace{-.7cm}
\end{figure}

Figure \ref{fig8} presents how the load balancing mechanism distributed the load among the different instances.\ In addition, it shows the circuit breaker behaviour regarding different errors to improve the reliability of the system.

\vspace{-.6cm}

The client-side strategy load balancing automatically distributed concurrent calls to the available Backserver instances.\ The Netflix Ribbon load balancer continuously rotated a list of Backserver instances that were attached to it (the Round Robin method).\ In addition, to manage failures that happened in a service (e.g., timeout), Hystrix prevented cascading failures to other services, which improved the fault tolerance of our system.\ Broken service instances automatically recovered and registered themselves into the Eureka service registry, which made the designed microservices recoverable.

\begin{tcolorbox}
\textbf{Summary.} Although failures often happen in services, load balancing mechanisms were able to skip unhealthy instances.
\end{tcolorbox}

\vspace{-.7cm}
\section{Discussions}
\label{VII}
\vspace{-.3cm}

The objective of this research was to review whether the recent architectural design styles, such as microservices, could address CD and DevOps in the automotive domain.\ In an experimental setting, we evaluated how quality attributes such as the scalability and reliability of services could be improved by microservices patterns.

\vspace{-.5cm}
\subsubsection{RQ1.\ Can the microservices architecture design enable over-the-air continuous delivery from cloud platforms in the automotive domain?}
A previous study \cite{Chen-microservice} noted that to maintain continuous software delivery, it is necessary to address architectural challenges, such as the deployability and modifiability.\ Our findings showed that the proposed architecture could improve the deployability of the system as there was no need to resolve the conflicts between changes afterwards.\ Furthermore, we could deploy changes in different services independently and quickly without any interruption in other services.\ 

We noticed that microservices created the possibility to make the changes localised to one service while other services were not affected.\ We had lightweight services that made any update in the codes easier.\ In safety-critical systems, such as ACPs, it is vital that changes in a service or technology do not interrupt other running services.\ Our findings showed that microservices could improve the modifiability of the architecture.\ Although the designed architecture could enable the CD in this domain by sending OTA messgaes to the rover, there were several failed and timeout calls that should be optimised with respect to different service level agreements.

\vspace{-.5cm}
\subsubsection{RQ2.\ How can the microservices architecture design improve the reliability and scalability of services in cloud platforms in the automotive domain?}
Scalability is the property of a system that handles a growing amount of requests by adding resources to the system.\ The Backserver instances allowed us to support a good number of concurrent calls coming from the Client to the rover.\ The Backserver was also stateless, which did not retain consumer states.\ It enabled us to have autoscaling of the services when the load required.\ The load balancing mechanism in our system could also distribute the load automatically among available service instances.

Our findings in this study showed how the fault-tolerant mechanisms, such as the circuit breaker, could handle the resiliency and reliability in our proposed architecture.\ We defined different thresholds such as the error threshold percentage and request volume threshold to force the circuit breaker to open and prevent slow or failed calls from interrupting other services in our architecture, which improved reliability of the architecture.

\vspace{-.55cm}
\subsection{Threats to validity}
\vspace{-.2cm}

Construct validity, in our research, is concerned with using the right measures in our experiment.\ To assess the reliability and scalability, we used the common metrics that are widely applied in the literature (see \cite{bass2015devops,newman2015building}).\ Internal validity concerns the relationship between the constructs and the proposed explanation.\ Our implementation was run in three scenarios in a laboratory experimental setting with specific and defined objectives.\ Although we established a controlled environment, aspects related to the performance of Azure cloud platform or 4G network connection could not be customised or controlled.\ In addition, the implementation and results were discussed and reviewed among the authors of this study.\ In our experiment, we selected the technology stacks that are commonly used by companies and the performance analysis of those technologies are out of scope of this research.

External validity is related to the generalisability of the study.\ A previous study \cite{Aderaldo} noted that it is not essential to satisfy all requirements by a given benchmark candidate to be considered useful for empirical research.\ We applied microservices patterns from scientific literature, established a controlled experiment with three defined scenarios, and used a real-world project to evaluate the behaviour of one single microservice in the designed architecture.\ Future studies can replicate the experiment with multiple services in real continuous software delivery environments in the automotive domain to evaluate generalisability of the results.\ Reliability concerns the repeatability of the research procedure and conclusions.\ We explained in detail the experimental setting and all publicly available materials, which can be applied by future studies.

\vspace{-.5cm}
\section{Conclusion}
\label{VIII}
\vspace{-.35cm}

Automotive cloud platforms have received increasing attention from research and industrial communities.\ To increase the reliability and scalability in ACPs and enable continuous software delivery in the automotive domain, we proposed a microservices architecture for a real-world project called Eclipse Kuksa and ran an experiment to evaluate the designed architecture.

Our findings showed that the proposed architecture could handle CD through improving the deployability, modifiability, and availability of the architecture.\ Our designed architecture could address quality issues, such as payload distribution among different instances and the resiliency of services.\ The research findings showed that microservices are an interesting design alternative to address quality concerns of future cloud platforms in the automotive domain.


\vspace{-.4cm}
\begin{thebibliography}{8}
\vspace{-.3cm}

\bibitem{Aderaldo}
Aderaldo C.\ M., Mendon\c{c}a N.C., Pahl C., Jamshidi P.: Benchmark requirements for microservices architecture research.\ In: 1st Int.\ Workshop on Establishing the Community-Wide Infrastructure for Arch.-Based Soft\ Eng., pp.\ 8-13.\ IEEE (2017).

\bibitem{balalaie2016microservices}
Balalaie, A., Heydarnoori, A., Jamshidi, P.: Microservices architecture enables devops: Migration to a cloud-native architecture.\ IEEE Software \textbf{33}, 42-52 (2016).

\bibitem{balalaie2015migrating}
Balalaie, A., Heydarnoori, A., Jamshidi, P.: Migrating to cloud-native architectures using microservices: An experience report.\ In: Adv.\ in Service-Oriented and Cloud Comp., pp.\ 201-215.\ Springer, Switzerland (2015).

\bibitem{balalaie2018microservices}
Balalaie, A., Heydarnoori, A., Jamshidi, P., Tamburri, D.A., Lynn, T.: Microservices migration patterns.\ Software: Practice and Experience.\ J.\ Software: Prac.\ and Expe.\ \textbf{48}, pp.\ 2019-2042 (2018).

\bibitem{bass2015devops}
Bass, L., Weber, I., Zhu, L.: DevOps: A software architect's perspective.\ Addison-Wesley Professional (2015).

\bibitem{Chen-microservice}
Chen, L.: Microservices: Architecting for continuous delivery and DevOps.\ In: Int.\ Conf.\ on Software Arch.\ (ICSA), pp.\ 39-397.\ IEEE (2018).

\bibitem{contreras2017internet}
Contreras-Castillo, J., Zeadally, S., Guerrero-Ibanez, J.A.: Internet of vehicles: Architecture, protocols, and security.\ Internet of Things J.\ \textbf{5}, pp.\ 3701-3709 (2018).

\bibitem{datta2015onem2m}
Datta, S.\ K., Gyrard, A., Bonnet, C., Boudaoud, K.: oneM2M architecture based user centric IoT application development.\ In: 3rd Int.\ Conf.\ on Future Internet of Things and Cloud, pp.\ 100-107.\ IEEE (2015).

\bibitem{dragoni2017microservices}
Dragoni, N., Dustdar, S., Larsen, S.T., Mazzara, M.: Microservices: Migration of a mission critical system.\ arXiv preprint arXiv:1704.04173 (2017).

\bibitem{ebert2017automoitve} 
Ebert, C., Favaro, J.: Automotive software.\ IEEE Software \textbf{34}, pp.\ 33-39 (2017).

\bibitem{ebert2016devops}
Ebert, C., Gallardo, G., Hernantes, J., Serrano, N.: Devops.\ IEEE Software \textbf{33}, 94-100 (2016).

\bibitem{Fiosina-bigdata}
Fiosina, J., Fiosins, M., M{\"u}ller, J.P.: Big data processing and mining for next generation intelligent transportation systems.\ J.\ Teknologi \textbf{63}, pp.\ 21-38 (2013).

\bibitem{MartinFowler}
Fowler, M., Lewis, J.: Microservices.\ https://martinfowler.com/articles/microservices.html.

\bibitem{googlecloud}
Google Cloud: Designing a Connected Vehicle Platform on Cloud IoT Core - 2019-05-07.\ https://cloud.google.com/solutions/designing-connected-vehicle-platform.

\bibitem{haberle2015connected}
H\"aberle, T., Charissis, L., Fehling, C., Nahm, J., Leymann, F.: The connected car in the cloud: A platform for prototyping telematics services.\ IEEE Software \textbf{32}, 11-17 (2015).

\bibitem{Haghighatkhah}
Haghighatkhah A., Banijamali A., Pakanen O., Oivo M., Kuvaja P.: Automotive software engineering: A systematic mapping study. J. Syst. Soft. 128, 25-55 (2017).

\bibitem{he2014developing}
He, W., Yan, G., Da, Xu L.: Developing vehicular data cloud services in the IoT environment.\ IEEE Trans.\ on Ind.\ Info.\ \textbf{10}, pp.\ 1587-1595 (2014).

\bibitem{shiftmobility}
Jain, P.: Automotive Cloud Technology to Drive Industry’s New Business Models - 2019-05-07.\ http://shiftmobility.com/2017/06/automotive-cloud-technology-drive-automotive-industrys-new-business-models.

\bibitem{armbrust2010view}
Armbrust, M., Fox, A., Griffith, R., Joseph, A.\ D., Katz, R., Konwinski, A., Lee, G., Patterson, D., Rabkin, A., Stoica, A., Zaharia, M.: A view of cloud computing.\ Commun of the ACM \textbf{53} (2010).

\bibitem{levcovitz2016towards}
Levcovitz, A., Terra, R., Valente, M.T.: Towards a technique for extracting microservices from monolithic enterprise systems.\ arXiv:1605.03175, (2016).

\bibitem{lu2014connected}
Lu, N., Cheng, N., Zhang, N., Shen, X., Mark, J.W.: Connected vehicles: Solutions and challenges.\ Internet of Things J.\ \textbf{1}, pp.\ 289-299.\ IEEE (2014).

\bibitem{mietzner2011horizontal}
Mietzner, R., Leymann, F., Unger, T.: Horizontal and vertical combination of multi-tenancy patterns in service-oriented applications.\ Enterprise Info.\ Syst.\ \textbf{5}, pp.\ 59-77 (2011).

\bibitem{newman2015building}
Newman, S.: Building microservices: designing fine-grained systems.\ O'Reilly Media, Inc.\ (2015).

\bibitem{o2007quality}
O'Brien, L., Merson, P., Bass, L.: Quality attributes for service-oriented architectures.\ In: Proc.\ of the Int.\ Workshop on Syst.\ Dev.\ in SOA Env., pp.\ 3 (2007).

\bibitem{pahl2016microservices}
Pahl, C., Jamshidi, P.: Microservices: A systematic mapping study.\ In: Proc.\ of the 6th Int.\ Conf.\ on Cloud Computing and Services Science, pp.\ 137-146 (2016).

\bibitem{rufino2017monitoring}
Rufino, J., Alam, M., Ferreira, J.: Monitoring V2X applications using DevOps and docker.\ In: Int.\ Smart Cities Conf., pp.\ 1-5 (2017).

\bibitem{Serrano-realtime}
Serrano, D., Baldassarre, T., Stroulia, E.: Real-time traffic-based routing, based on open data and open-source software.\ In: 3rd World Forum on Internet of Things, pp.\ 661-665 (2016).

\bibitem{shavit2007firmware}
Shavit, M., Gryc, A., Miucic, R.: Firmware update over the air (FOTA) for automotive industry.\ SAE Tech.\ (2007).

\bibitem{stol2018abc}
Stol, K., Fitzgerald, B.: The ABC of software engineering research.\ ACM Trans.\ on Software Eng.\ and Meth.\ \textbf{27}, 11 (2018).

\bibitem{taibi2018architectural}
Taibi, D., Lenarduzzi, V., Pahl, C.: Architectural patterns for microservices: A systematic mapping study.\ In: Proc.\ of the 8th Int.\ Conf.\ on Cloud Computing and Services Science, pp.\ 221-232 (2018).

\bibitem{thones2015microservices}
Th{\"o}nes, J.: Microservices.\ IEEE Software \textbf{32}, pp.\ 116-116 (2015).

\bibitem{yang-intelligent}
Yang, M., Mahmood, M., Zhou, X., Shafaq, S., Zahid, L.: Design and implementation of cloud platform for intelligent logistics in the trend of intellectualization.\ China Commu.\ \textbf{14}, pp.\ 180-191 (2017).

\bibitem{zeller2013towards}
Zeller, M., Prehofer, C., Krefft, D., Weiss, G.: Towards runtime adaptation in AUTOSAR.\ In: 5th Workshop on Adaptive and Reconfigurable Embedded Syst.\ \textbf{10}, pp.\ 17-20 (2013).

\bibitem{zhang2014defending}
Zhang, T., Antunes, H., Aggarwal, S.: Defending connected vehicles against malware: Challenges and a solution framework.\ Internet of Things J.\ \textbf{1}, pp.\ 10-21.(2014).

\bibitem{zhu2016devops}
Zhu, L., Bass, L., Champlin-Scharff, G.: DevOps and its practices.\ IEEE Software \textbf{33}, 32-34 (2016).

\end{thebibliography}
\end{document}